\documentclass{SciPost}

\hypersetup{
    colorlinks,
    linkcolor={red!50!black},
    citecolor={blue!50!black},
    urlcolor={blue!80!black}
}

\usepackage[bitstream-charter]{mathdesign}
\DeclareSymbolFont{usualmathcal}{OMS}{cmsy}{m}{n}
\DeclareSymbolFontAlphabet{\mathcal}{usualmathcal}

\fancypagestyle{SPstyle}{
\fancyhf{}
\lhead{\colorbox{scipostblue}{\bf \color{white} ~SciPost Physics Core }}
\rhead{{\bf \color{scipostdeepblue} ~Submission }}

\fancyfoot[C]{\textbf{\thepage}}
}
\usepackage{scalerel}
\usepackage{soul}
\usepackage{tikz}
\usetikzlibrary{calc}
\usetikzlibrary{patterns}
\usetikzlibrary{svg.path}
\definecolor{orcidlogocol}{HTML}{A6CE39}

\tikzset{
  orcidlogo/.pic={
    \fill[orcidlogocol] svg{M256,128c0,70.7-57.3,128-128,128C57.3,256,0,198.7,0,128C0,57.3,57.3,0,128,0C198.7,0,256,57.3,256,128z};
    \fill[white] svg{M86.3,186.2H70.9V79.1h15.4v48.4V186.2z}
                 svg{M108.9,79.1h41.6c39.6,0,57,28.3,57,53.6c0,27.5-21.5,53.6-56.8,53.6h-41.8V79.1z M124.3,172.4h24.5c34.9,0,42.9-26.5,42.9-39.7c0-21.5-13.7-39.7-43.7-39.7h-23.7V172.4z}
                 svg{M88.7,56.8c0,5.5-4.5,10.1-10.1,10.1c-5.6,0-10.1-4.6-10.1-10.1c0-5.6,4.5-10.1,10.1-10.1C84.2,46.7,88.7,51.3,88.7,56.8z};
  }
}

\newcommand\orcid[1]{\!%
  \href{https://orcid.org/#1}{%
    \mbox{%
      \scaleto{%
        \begin{tikzpicture}[yscale=-1,transform shape]
          \pic{orcidlogo};
        \end{tikzpicture}
      }{8pt}%
    }%
  }%
}
\begin{document}

\pagestyle{SPstyle}

\begin{center}{\Large \textbf{\color{scipostdeepblue}{
Scaling of free cumulants in closed system-bath setups
}}}\end{center}

\begin{center}\textbf{
Merlin F\"{u}llgraf~\orcid{0009-0000-0409-5172} 
\textsuperscript{$\star$},
Jochen Gemmer~\orcid{0000-0002-4264-8548}  and
Jiaozi Wang~\orcid{0000-0001-6308-1950} \textsuperscript{$\dagger$}
}\end{center}

\begin{center}
University of Osnabr\"{u}ck, Institute of Physics, D-49076 Osnabr\"{u}ck, Germany
\\

$\star$ \href{merlin.fuellgraf@uos.de}{\small merlin.fuellgraf@uos.de}\,,\quad
$\dagger$ \href{jiaowang@uos.de}{\small jiaowang@uos.de}
\end{center}

\section*{\color{scipostdeepblue}{Abstract}}
\textbf{\boldmath{%
The Eigenstate Thermalization Hypothesis (ETH) has been established as a cornerstone for understanding thermalization in quantum many-body systems. 
Recently, there has been growing interest in the full ETH, which extends the framework of the conventional ETH and postulates a smooth function to describe the multi-point correlations among matrix elements.
Within this framework, free cumulants play a central role, and most previous studies have primarily focused on closed systems. In this paper, we extend the analysis to a system–bath setup, considering both an idealized case with a random-matrix bath and a more realistic scenario where the bath is modeled as a defect Ising chain. In both cases, we uncover a universal scaling of the microcanonical free cumulants of observables associated with the central system Hamiltonian with respect to the interaction strength. Furthermore we establish a connection between this scaling behavior and the thermalization dynamics of the thermal free cumulants of corresponding observables.}
}

\vspace{\baselineskip}




\vspace{10pt}
\noindent\rule{\textwidth}{1pt}
\tableofcontents
\noindent\rule{\textwidth}{1pt}
\vspace{10pt}

\section{Introduction}
\label{sec:intro}
Thermalization of quantum many-body systems is one of the central questions in non-equilibrium statistical mechanics. 
Among the various frameworks that have been developed, the Eigenstate Thermalization Hypothesis \cite{deutsch91,Srednicki94} (ETH) is one among the most promising, which receive lots of attention during the last decade.

The ETH explains the eventual thermalization of the system by postulating a universal structure for the matrix elements of physical observables in the energy eigenbasis, known as the ETH ansatz\cite{Srednicki1999,Srednicki1996,Review-ETH}. Despite the lack of a rigorous proof,  the ETH ansatz has been verified numerically in numerous studies \cite{rigol08,eth-num-PhysRevB.99.155130,eth-num-PhysRevE.100.062134,eth-num-PhysRevE.82.031130,eth-num-PhysRevE.87.012118,eth-num-PhysRevE.89.042112,eth-num-PhysRevE.90.052105,eth-num-PhysRevE.91.012144,eth-num-PhysRevE.93.032104,eth-num-PhysRevE.96.012157,Delacretaz:2022ojg,PhysRevB.103.235137-eth-Vidmar,PhysRevB.110.104202-WETH-Vidmar24} and is now widely believed to hold in generic non-integrable systems for few-body observables.

However, ETH in its conventional form (the ETH ansatz) is not the end of the story.
Recently, several studies have focused on correlations among the matrix elements \cite{foini2019eigenstate, chan2019eigenstate, murthy2019bounds, richter2020eigenstate, brenes2021out, wang2021eigenstate, dymarsky2022bound,PhysRevX.14.031029-eth-correlation-Luitz}, and a generalized form of the ETH has been proposed \cite{FoiniPRE-geth}, often referred to as the full ETH. The full ETH extends the standard framework by incorporating multi-point correlations, which are described in terms of smooth functions.
Furthermore, the full ETH has intriguing connections to free probability theory\cite{PhysRevLett.129.170603-geth-Silvia}, providing a powerful mathematical perspective on many-body thermalization. 
In recent years, this extended framework has attracted growing attention \cite{wang2025eigenstatethermalizationhypothesiscorrelations,PhysRevLett.134.140404-geth,PhysRevLett.129.170603-geth-Silvia,foini2019eigenstate2,Fava2025Designs, PhysRevB.111.054303-silvia-geth,fritzsch2025freecumulantseigenstatethermalization-geth,alves2025probeseigenstatethermalizationergodicitybreaking-geth,PhysRevX.13.031033-geth-Sonner,PhysRevD.108.066015-geth-prd-Sonner}. However, most of the existing studies focused primarily on isolated quantum systems.

In this paper, we extend the investigation of the full ETH to open-system setups by considering a central system coupled to a quantum chaotic bath. 
We study the structure of smooth multi-point correlation functions in this setting, with particular emphasis on the energy scale within which such correlation functions become structureless.
Using large-scale numerical simulations, we demonstrate the existence of such an energy scale and identify its scaling behavior with respect to the interaction strength and system size. Furthermore, we investigate the dynamical significance of this energy scale by analyzing the thermal free cumulants of the corresponding observables.

The remainder of the paper is organized as follows. Section\ \ref{sec-framework} introduces the main framework, including the setup, the Full ETH and the definition of free cumulants. Section\ \ref{sec_numerics} presents the numerical results: Subsection\ \ref{subsec_rm} focuses on a random-matrix model, where we first describe the setup and then discuss microcanonical and thermal free cumulants. Subsection\ \ref{subsec_ising} follows a similar structure but considers a defect Ising chain as a bath. Finally, Section\ \ref{sec_conclusion} provides the conclusions and outlook.

\section{Framework\label{sec-framework}}

\subsection{Model and observables}
We consider an open-system setup, where the total Hamiltonian can, in full generality, be written as
\begin{align}
    \mathcal{H}=\mathcal{H}_{S}+\lambda\mathcal{H}_{I}+\mathcal{H}_{B}.\label{eq_gen-hamiltonian}
\end{align}
Here $\mathcal{H}_S$ refers to the Hamiltonian of a comparatively small system,  $\mathcal{H}_B$ represents a bath and $\mathcal{H}_I$ describes the interaction between the two, moderated by the coupling strength parameter $\lambda$. The eigenstates and eigenvalues of $\mathcal{H}$ are denoted by $|i\rangle$ and $E_i$, respectively. We consider operators of the system, generically denoted by $A$.

\subsection{Full Eigenstate Thermalization Hypothesis}
The statistical properties of the matrix elements of an operator $A$ in the energy eigenbasis, $A_{ij} = \langle i | A | j \rangle$, can be described by the full ETH framework, which extends the conventional ETH to include multi-point correlations.  Specifically, it states that the statistical averages of products of matrix elements with repeated indices factorize, while with distinct indices,
$i_1 \ne i_2 \ne \dots \ne i_n$ satisfy
\begin{align}
\label{eth1}
\overline{A_{i_1 i_2}A_{i_2 i_3}\dots A_{i_n i_1}} = \Omega_{E^+}^{1-n}\, & F_{e^+}^{(n)} \left(\omega_{i_1 i_2},\dots, \omega_{i_{n-1} i_n}\right)\ ,
\end{align}
where $\Omega_{E^+}$ is the density of states at the average energy $E^+ \equiv (E_{i_1}+\cdots + E_{i_n})/n$. Here the $F_{e^+}^{(n)}$ are \emph{smooth functions} of the average energy density $e^+ = E^+/L $ and the eigenenergy differences $\omega_{i_1 i_2}=E_{i_1}-E_{i_2}$ \footnote{For $n=1,2$, one recovers the conventional ETH \cite{Srednicki1999}, where $F_e^{(1)} = \mathcal A(e)$ is the constant equilibrium average and $F_e^{(2)}(\omega)=|f(e, \omega)|^2$ is the dynamical correlation function.}. 
These smooth functions $F_{e^{+}}^{(n)}(\vec{\omega})$, with $\vec{\omega} = (\omega_{i_1 i_2},\omega_{i_2 i_3}, \ldots, \omega_{i_{n-1},i_{n}})$, will be referred to as ETH functions throughout the rest of the paper. 

For $n=2$, the full ETH ansatz reduces to the conventional ETH ansatz \cite{Srednicki1996,Srednicki1999} and the ETH function $F_{e^{+}}^{(2)}(\omega)$ are well understood in this case. It has been found that, for generic observables in chaotic systems, there often exists an energy scale $\Delta E_{\text{eq}}^{(2)}$ within which $F_{e^{+}}^{(2)}(\omega)$ is a constant \cite{PhysRevB.96.104201-Papic-Thouless,PhysRevE.102.062113-thouless}, i.e.,
\begin{equation}
    F_{e^{+}}^{(2)}(\omega)=\text{const.},\ \forall|\omega|\le\Delta E_{\text{eq}}^{(2)}\ .
\end{equation}
The energy scale $\Delta E_{\text{eq}}^{(2)}$ is often referred to as Thouless energy. 

A natural question is, whether a corresponding energy scale $\Delta E_{\text{eq}}^{(n)}$ exists for more generic $n$ such that
\begin{equation}\label{eq_cum_powerlaw0}
F_{e^{+}}^{(n)}(\vec{\omega})=\text{const.},\ \ \ \forall\ |\omega_{i_{1}i_{2}}|\le\Delta E_{\text{eq}}^{(n)},\ldots,|\omega_{i_{n-1}i_{n}}|\le\Delta E_{\text{eq}}^{(n)}\ .
\end{equation}
Previous studies mainly focused on closed-system scenarios and evidence for the existence of $\Delta E_{\text{eq}}^{(n)}$ for general few-body observables in chaotic quantum many-body systems has been presented \cite{wang-unitary-symmetry}. In addition, the values of $\Delta E_{\text{eq}}^{(n)}$ are found to be strongly dependent on the observable, the system size, and, in most cases, also on $n$.
So far, not much has been done in open-system scenarios.
Most questions remain open, for instance, whether such an energy scale $\Delta E_{\text{eq}}^{(n)}$ exists, and if so, how it scales with parameters such as the coupling strength and the bath sizes, as well as what impact it has on the system’s dynamics. 
These will be the main questions explored in this paper.

\subsection{ETH free cumulants}
To understand the impact of the full ETH on the dynamics of the corresponding observable, let us consider the $n$-time thermal correlation function
\begin{align}
    C^\beta_{n}(\vec{t})= & \frac{1}{Z}\text{Tr}[A(t_{1})e^{-\frac{\beta H}{n}}A(t_{2})e^{-\frac{\beta H}{n}}\cdots e^{-\frac{\beta H}{n}}A(t_{n})e^{-\frac{\beta H}{n}}] \nonumber \\
     & =\frac{1}{Z}\left(\sum_{i_{1},\ldots,i_{n}}e^{-\frac{\beta}{n}(E_{i_{1}}+\cdots+E_{i_{n}})}A_{i_{1}i_{2}}A_{i_{2}i_{3}}\cdots A_{i_{n}i_{1}}e^{iE_{i_{1}}(t_{1}-t_{n})+\cdots+iE_{i_{n}}(t_{n}-t_{n-1})}\right)\label{eq-Cnt}
\end{align}
where $\vec{t} = (t_1,t_2,\ldots,t_n)$ and $Z=\text{Tr}[e^{-\beta H}]$, indicating the partition function. When the sum of Eq.~\eqref{eq-Cnt} is only taken over distinct indices, one can define the ETH free cumulants, 
\begin{equation}
\kappa_{n}^{\text{ETH}}(\vec{t})=\frac{1}{Z}\left(\sum_{\substack{i_{1},\ldots,i_{n}\\
i_{1}\neq i_{2}\cdots\neq i_{n}
}
}e^{-\frac{\beta}{n}(E_{i_{1}}+\cdots+E_{i_{n}})}A_{i_{1}i_{2}}A_{i_{2}i_{3}}\cdots A_{i_{n}i_{1}}e^{iE_{i_{1}}(t_{1}-t_{n})+\cdots+iE_{i_{n}}(t_{n}-t_{n-1})}\right) .
\end{equation}
It has been shown that, in the thermodynamic limit,  $\kappa_{n}^{\text{ETH}}(\vec{t})$ is related to the ETH function $F^{(n)}_{e^+}$ through 
\begin{equation}\label{eq-kETH0}
\kappa_{n}^{\text{ETH}}(\vec{t})=\int d\vec{\omega}F_{e_{\beta}}^{(n)}(\vec{\omega})e^{i\vec{\omega}\cdot\vec{t}-\beta\vec{\omega}\cdot\vec{\ell}_{n}},
\end{equation}
where $\vec{\ell}_{n}=(\frac{n-1}{n},\frac{n-2}{n},\ldots,\frac{1}{n})$ and $e_{\beta}=\frac{1}{ZL}\text{Tr}[e^{-\beta H}H]$. 
Applying the Fourier transform on both sides of Eq.~\eqref{eq-kETH0}, one obtains
\begin{equation}\label{eq-komega}
    \kappa_{n}^{\text{ETH}}(\vec{\omega})\equiv\text{FT}[\kappa_{n}^{\text{ETH}}(\vec{t})]=F_{e_{\beta}}^{(n)}(\vec{\omega})e^{-\beta\vec{\omega}\cdot\vec{\ell}_{n}} .
\end{equation}


\subsection{Thermal free cumulants\label{subsec-therm-cumulants}}
After introducing the ETH free cumulants, we now turn to a closely related quantity: the thermal free cumulants. The thermal free cumulants $\kappa_{n}^\beta(t)$ are routinely defined recursively via
\begin{align}\label{eq_def_cum}
C_n^{\beta}(t_{1},t_{2},\ldots,t_{n})=\sum_{\pi\in NC(n)}\kappa^\beta_{\pi}(t_{1},t_{2},\ldots,t_{n}),
\end{align}
where the sum runs over all \textit{non-crossing} partitions $NC(n)$ of the set given by $\{1,\dots,n\}$.  See Ref.\ \cite{PhysRevLett.129.170603-geth-Silvia} for more details. For instance, in case of $C_{1}^{\beta}(0)=0$ 
\begin{align}
    C_{1}^{\beta}(0) & =\kappa_{1}^{\beta}(0)=0 \\
    C_{2}^{\beta}(t_{1},0) & =\kappa_{2}^{\beta}(t_{1},0) \\
    C_{3}^{\beta}(t_{2},t_{1},0) & =\kappa_{3}^{\beta}(t_{2},t_{1},0) \\
    C_{4}^{\beta}(t_{3},t_{2},t_{1},0) & =\kappa_{4}^{\beta}(t_{3},t_{2},t_{1},0)+\kappa_{2}^\beta(t_{3},t_{2})\kappa_{2}^\beta(t_{1},0)+\kappa_{2}^\beta(t_{3},0)\kappa_{2}^\beta(t_{2},t_{1}) \\
    \ldots & = \ldots
\end{align}
Eq.\ (\ref{eq_def_cum}) can be easily inverted and the thermal free cumulants $\kappa^{\beta}_{n}(\vec{t})$ can also be expressed in terms of the correlation functions $C^{\beta}_{n}(\vec{t})$. In other words, the $\kappa^{\beta}_{n}(\vec{t})$ are unambiguously fixed by $C^{\beta}_{q}(\vec{t})$ ($q\le n$).

The full ETH implies that  thermal free cumulants $\kappa^{\beta}_{n}(\vec{t})$ are related to ETH free cumulants  $\kappa^{\beta}_{n}(\vec{t})$ as \cite{PhysRevLett.129.170603-geth-Silvia}
\begin{equation}
    \kappa_{n}^{\beta}(\vec{t})=\kappa_{n}^{\text{ETH}}(\vec{t})+{\cal O}(L^{-1})\ .
\end{equation}
In the thermodynamic limit $L \rightarrow \infty$, both quantities coincide $\kappa_{n}^{\beta}(\vec{t})=\kappa_{n}^{\text{ETH}}(\vec{t})$.
In this case, a connection between thermal free cumulants $\kappa_{n}^{\beta}(\vec{t})$ and the ETH function $F_{e^{+}}^{(n)}(\vec{\omega})$ can be established using Eq.~\eqref{eq-komega},
\begin{equation}\label{eq-kbetaomega}
\kappa_{n}^{\beta}(\vec{\omega})\equiv\text{FT}[k_{n}^{\beta}(\vec{t})]=F_{e_{\beta}}^{(n)}(\vec{\omega})e^{-\beta\vec{\omega}\cdot\vec{\ell}_{n}}.    
\end{equation}
From here on, we restrict ourselves to the case of infinite temperature ($\beta = 0$), but the consideration can in principle be generalized to the finite-temperature case.
At $\beta = 0$, Eq.~\eqref{eq-kbetaomega} becomes
\begin{equation}\label{eq-komega-2}
    \kappa_{n}(\vec{\omega})=F_{e_{0}}^{(n)}(\vec{\omega}) ,
\end{equation}
where we omit the superscript $\beta = 0$ for brevity.
It is clear from Eq.~\eqref{eq-komega-2} that the non-trivial dependence of the ETH function $F_{e_{0}}^{(n)}(\vec{\omega})$ governs the out-of-equilibrium dynamics of the thermal free cumulants $\kappa_n(\vec{t})$.
At long times $t$, the low-frequency structure of $F_{e_{0}}^{(n)}(\vec{\omega})$ becomes most relevant.

In chaotic systems, $\kappa_{n}(\vec{t})$ generally thermalizes after a characteristic timescale $T_{\text{eq}}^{(n)}$. Based on the preceding discussion, a connection can be drawn between the thermalization time $T_{\text{eq}}^{(n)}$ and the energy scale $\Delta E_{\text{eq}}^{(n)}$, which characterizes the energy window within which $F_{e_{0}}^{(n)}(\vec{\omega})$ becomes structureless:
\begin{equation}\label{eq-ET}
    \Delta E_{\text{eq}}^{(n)} \sim \frac{2\pi}{T_{\text{eq}}^{(n)}}\ .
\end{equation}
This relation highlights the dynamical significance of the energy scale $\Delta E_{\text{eq}}^{(n)}$. While for the second-order cumulants this thermalization time characterizes the timescales of the system in the linear response regime, i.e.\ for near-equilibrium initial states, such thermalization time as in Eq.\ (\ref{eq-ET}) are more relevant for far-from-equilibrium states\ \cite{PhysRevE.99.050104-jonas}. 
Further, these thermalization times are in close relation to freeness timescales, as will be discussed in Subsection \ref{subsec_rm}.

For $n=2$, the thermal free cumulants $\kappa_2(t)$, which coincides with the connected autocorrelation function, have been thoroughly studied in the open-system framework \cite{breuer2002book-qop,weiss2012book-qop}, particularly in the weak-coupling regime ($\lambda \rightarrow 0$) where the Markovian approximation can be applied. For example, considering the system Hamiltonian as an observable,
it can be shown that, in the Markovian regime
\begin{equation}
    \kappa_{2}(t)\propto e^{-\Gamma t},\ \ \text{for}\ L\rightarrow \infty\  ,
\end{equation}
where $\Gamma \propto \lambda^2$. Correspondingly, using Eq.~\eqref{eq-kbetaomega}, the ETH function $F^{(2)}_{e_0}(\omega)$ takes the form 
\begin{equation}
    F^{(2)}_{e_0}(\omega)\propto\frac{1}{\omega^{2}+\Gamma^{2}} \  .
\end{equation}
The function $  F^{(2)}_{e_0}(\omega)$ can be regarded as constant at low frequencies $\omega \ll \Gamma \propto \lambda^{2}$, thus it is expected that $\Delta E_{\text{eq}}^{(2)}\propto\Gamma\propto\lambda^{2}$.
The standard open-system framework cannot be straightforwardly generalized to the study of higher cumulants, and the general properties of $\kappa_{n}(\vec{t})$ and $F_{e_{0}}^{(n)}(\vec{\omega})$, as well as $\Delta E_{\text{eq}}^{(n)}$ remain unknown. The question we aim to address is whether the scaling 
$\Delta E_{\mathrm{eq}}^{(2)} \propto \lambda^{2}$, 
established for $n=2$ in the weak-coupling regime, 
also carries over to higher-order cumulants with $n>2$.

\subsection{Microcanonical free cumulants}
Studying the energy scale $\Delta E_{\text{eq}}^{(n)}$ by direct calculation of the ETH function $F_{e_{+}}^{(n)}(\vec{\omega})$ is very challenging, particularly for system sizes beyond the limits of exact diagonalization (ED). To this end, we employ an approach used in Ref.~\cite{wang-unitary-symmetry} and consider the microcanonically truncated operator
\begin{equation}
    A_{\Delta E}=P_{\Delta_E}A\ P_{\Delta E}, \quad\text{where} \ \ P_{\Delta E}=\sum_{\vert E_i-E_0\vert<\Delta E/2}\vert E_i\rangle\langle E_i\vert\label{eq_def_projector}\ .
\end{equation}
Here $E_0$ indicates the center of the energy window which is chosen as $E_0 = L e_0$, corresponding to the infinite temperature $\beta = 0$.
The microcanonical free cumulants are then given as a combination of the moments 
\begin{align}
    \mathcal{M}_n(\Delta E)=\frac{\text{tr}\left({A}_{\Delta E}\right)^n}{d_{\Delta E}}\label{eq_def_moments}
\end{align}
and read
\begin{align}
    \Delta_n(\Delta E)=\mathcal{M}_n-\sum_{j=1}^{n-1}\Delta_j \sum_{a_1+a_2+\dots+a_j=n-j}\mathcal{M}_{a_1}\dots\mathcal{M}_{a_j},
\end{align}
where $d_{\Delta E} = \text{Tr}[P_{\Delta E}]$.
In the eigenbasis $\Delta_n$ can be expressed as
\begin{equation}\label{eq-Delta0}
    \Delta_{n}(\Delta E)=\frac{1}{d_{\Delta E}}\sum_{\substack{i_{1},\ldots,i_{n},i_{1}\neq i_{2}\cdots\neq i_{n}\\
|E_{i_{k}}-E_{0}|<\frac{\Delta E}{2}\ \forall k
}
}A_{i_{1}i_{2}}A_{i_{2}i_{3}}\cdots A_{i_{n}i_{1}} .
\end{equation}
If the number of eigenstates within the energy windows is sufficiently large, one can 
replace $A_{i_{1}i_{2}}A_{i_{2}i_{3}}\cdots A_{i_{n}i_{1}}$ by its average 
$\overline{A_{i_{1}i_{2}}A_{i_{2}i_{3}}\cdots A_{i_{n}i_{1}}}$.
Inserting Eq.~\eqref{eth1}, Eq.~\eqref{eq-Delta0} becomes
\begin{equation}
\Delta_{n}(\Delta E)=\frac{1}{d_{\Delta E}}\sum_{\substack{i_{1},\ldots,i_{n},i_{1}\neq i_{2}\cdots\neq i_{n}\\
|E_{i_{k}}-E_{0}|<\frac{\Delta E}{2}\ \forall k
}
}F_{e^{+}}^{(n)}(\omega_{i_{1}i_{2}},\ldots,\omega_{i_{n-1}i_{n}}) .
\end{equation}
In the case of $\Delta E \le \Delta E^{(n)}_{\text{eq}}$, the condition $|\omega_{i_{k}i_{k+1}}| \le \Delta E^{(n)}_{\text{eq}}$ holds for all $k$ (with $i_{n+1} = i_1$). 
Recalling Eq.~\eqref{eq_cum_powerlaw0}, the ETH function $F_{e^{+}}^{(n)}(\vec{\omega})$ is structureless within this region,
$F_{e^{+}}^{(n)}(\vec{\omega})=\text{const.}$. As a result, one has 
\begin{equation}\label{eq-Delta1}
    \Delta_{n}(\Delta E)=\frac{1}{d_{\Delta E}}\sum_{\substack{i_{1},\ldots,i_{n},i_{1}\neq i_{2}\cdots\neq i_{n}\\
|E_{i_{k}}-E_{0}|<\frac{\Delta E}{2}\ \forall k
}
}1\simeq(d_{\Delta E})^{n-1} .
\end{equation}
For sufficiently small windows $\Delta E$, within which the density of states $\Omega_E$ can be regarded as constant, one has $d_{\Delta E} \propto \Delta E$, and Eq.~\eqref{eq-Delta1} becomes
\begin{equation}\label{eq_cum_powerlaw}
\Delta_{n}(\Delta E) \propto \Delta E^{n - 1}.
\end{equation}
Eq.~\eqref{eq_cum_powerlaw} will serve as our primary indicator for identifying the energy scale $\Delta E_{\text{eq}}^{(n)}$.

\section{Numerical investigation\label{sec_numerics}}
Below we expand on the models and observables considered in this work. Further details on the numerical methods may be found in App.\ \ref{app_num_details}.

\subsection{A random-matrix bath}\label{subsec_rm}
As a first model we consider a setup where a central spin is coupled to a random-matrix bath, also known as the spin-Gaussian orthogonal random matrices (GORM) model \cite{PhysRevE.68.066113-esposito}.
The Hamiltonian reads 
\begin{align}
    \mathcal{H}&=\omega_S\sigma_x^S+\lambda\sigma_z^S\otimes\mathcal{H}_{I}+\mathcal{H}_{B},\label{eq_ham_random}
\end{align}
where $\mathcal{H}_{I}$ and $\mathcal{H}_{B}$ are random matrices from the Gaussian orthogonal ensemble (GOE). Their entries are drawn from a Gaussian distribution with zero mean and variance $\sigma_0^2=\frac{1}{4d}$ with $d$ the dimension of the full system. To facilitate discussions we typically address the \textit{size} $L$ of the system, with $d=2^L$, rather than its dimension to treat it in the same language as other spin models. We fix the parameter $\omega_S$ to $\omega_S=0.05$, while varying the interaction strength $\lambda$. 

\begin{figure}[h]
    \centering
    \includegraphics[width=1.1\linewidth]{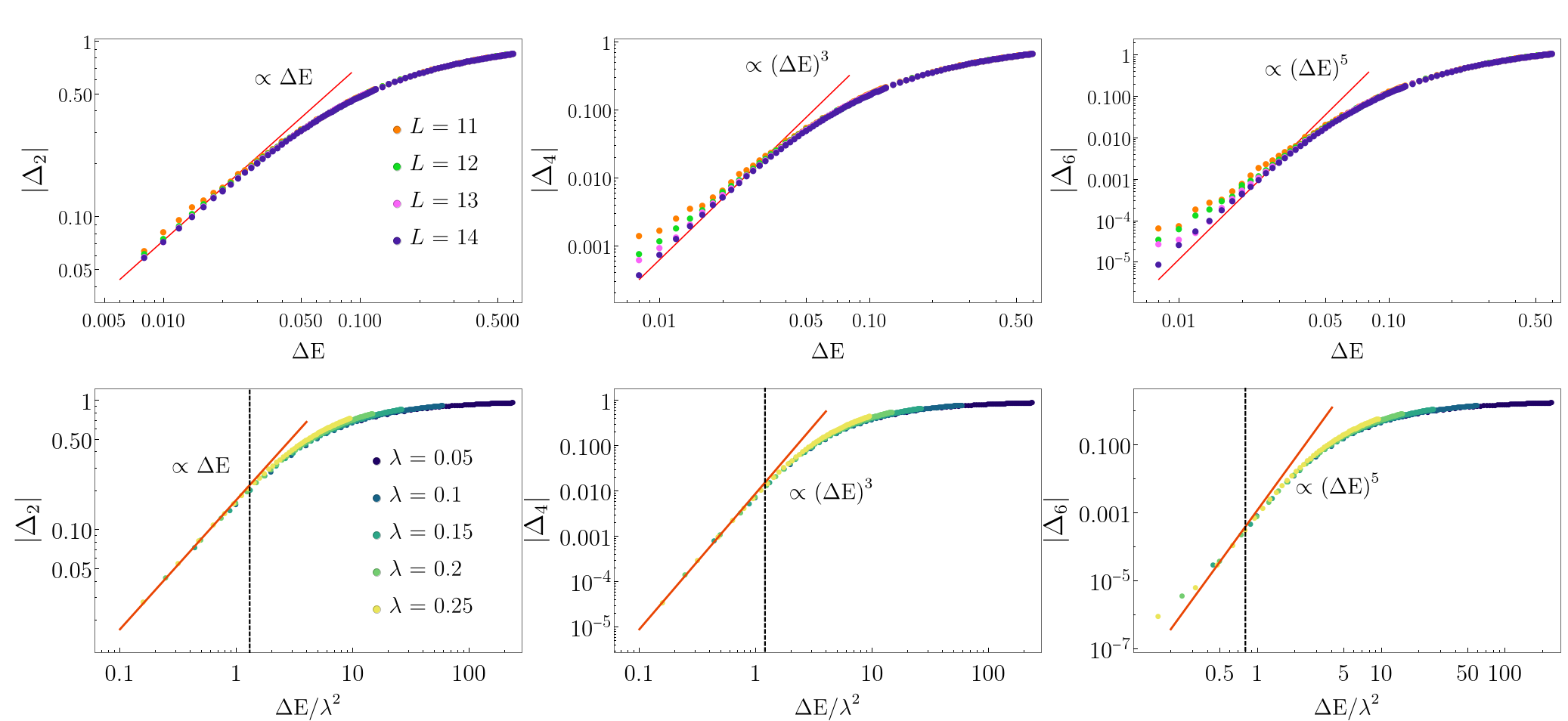}
    \caption{\textbf{Top:} Scaling of the microcanonical free cumulants $\Delta_n$ (with $n$ even) for the random-matrix model with interaction strength $\lambda=0.15$ for different system sizes. 
    \textbf{Bottom:} Scaling of the free cumulants $\Delta_n$ in the random-matrix model with size $L=14$ for different interaction strengths $\lambda=0.05,\dots,0.25$. Here the dashed black line indicates $\Delta E_U$ as a guide to the eye. Note that $L$ is related to the Hilbert space dimension of the whole system as $d = 2^L$.}
    \label{fig:cum_rm_even}
\end{figure}

The out-of-equilibrium dynamics of the central spin have been well studied, e.g.\ in Refs.\ \cite{PhysRevE.68.066113-esposito,PhysRevA.77.012108-qtherm,PhysRevA.90.022107-qtherm-qop,PhysRevLett.111.130408-qtherm-qop}, which can also be straightforwardly generalized to the study of the second thermal free cumulant of the system observables. 
In this model, there are three nontrivial system observables of interest, i.e., $\sigma^S_x$, $\sigma^S_y$ and $\sigma^S_z$.
In the weak-coupling limit $\lambda \rightarrow 0$, it can be derived that
\begin{equation}\label{eq-krm}
\begin{cases}
\kappa_{2}^{x}(t)\propto e^{-\Gamma_{R}t}\\
\kappa_{2}^{z}(t)\propto e^{-\Gamma_{R}t}\cos(2\omega t)\\
\kappa_{2}^{y}(t)\propto\frac{\partial^{2}\kappa_{2}^{z}(t)}{\partial t^{2}}
\end{cases},
\end{equation}
where  $\kappa^{x,y,z}_2(t)$ indicate the second thermal cumulants of $\sigma_{x,y,z}^{S}$ and $\Gamma_R \propto \lambda^2$. 

As an example, let us consider $A=\sigma_S^{x}$, which coincides with $\mathcal{H}_S$ up to a prefactor.
For weak system-bath coupling $\lambda$, according to Eq.~\eqref{eq-krm}, one has $F_{e_{0}}^{(2)}(\omega)\propto\frac{1}{\omega^{2}+\Gamma_{R}^{2}}$, which leads to $\Delta E_{\text{eq}}^{(2)}\propto\Gamma_{R}\propto\lambda^{2}$, see also the preceding discussion in Subsection\ \ref{subsec-therm-cumulants}. However, such scaling is not expected for $\sigma^S_{y,z}$, which is obvious from Eq.~\eqref{eq-krm}. In the main text, we focus on $\sigma^S_x$. In particular, we study numerically whether the scaling $\Delta E_{\text{eq}}^{(2)}\propto\lambda^{2}$ persists for higher-order cumulants, i.e.\ whether $\Delta E_{\text{eq}}^{(n)}\propto\lambda^{2}$ for $n > 2$, where we employ the criterion introduced in Eq.~\eqref{eq_cum_powerlaw} and analyze the 
microcanonical free cumulants.
In App.\ \ref{app_other_observable} we also investigate $\sigma_z^S$.

We first consider the scaling of the microcanonical free cumulants $\Delta_n$ with respect to the microcanonical energy window $\Delta E$ for different system-bath interaction strengths $\lambda$, see Fig.\ \ref{fig:cum_rm_even}.
We find that energy scales $\Delta E_{\text{eq}}^{(n)}$ exist, within which the microcanonical cumulants exhibit a power-law scaling with respect to $\Delta E$, as in Eq.\ (\ref{eq_cum_powerlaw}).
From the top panel, $\Delta E_{\text{eq}}^{(n)}$ appears to be almost independent of the system size. For small energy windows, deviations from Eq.~\eqref{eq_cum_powerlaw} can be observed, which are probably due to the finite size effect, since the validity of Eq.~\eqref{eq_cum_powerlaw} requires $d_{\Delta E} \propto 2^{L}
\Delta E\gg 1$.
However, it is evident that, the agreement with Eq.~\eqref{eq_cum_powerlaw} seems to extend to smaller $\Delta E$, for larger system sizes. It suggests that $\Delta_n$ will follow the power-law scaling indicated by Eq.~\eqref{eq_cum_powerlaw} to arbitrary small energy windows for $L\rightarrow \infty$.

From the lower panel, a scaling law 
$\Delta_{n}(\Delta E) = f\!\left(\frac{\Delta E}{\lambda^{2}}\right)$
with respect to the system--bath interaction strength $\lambda$ seems to hold for all cumulants considered. 
The observation reveals a scaling of the energy scale  $\Delta E^{(n)}_{\text{eq}}$ with  system–bath coupling strength $\Delta E_{\text{eq}}^{(n)}\propto\lambda^{2}$, which highlights the first main result of our paper.

\begin{figure}[h]
    \centering
    \includegraphics[width=1.1\linewidth]{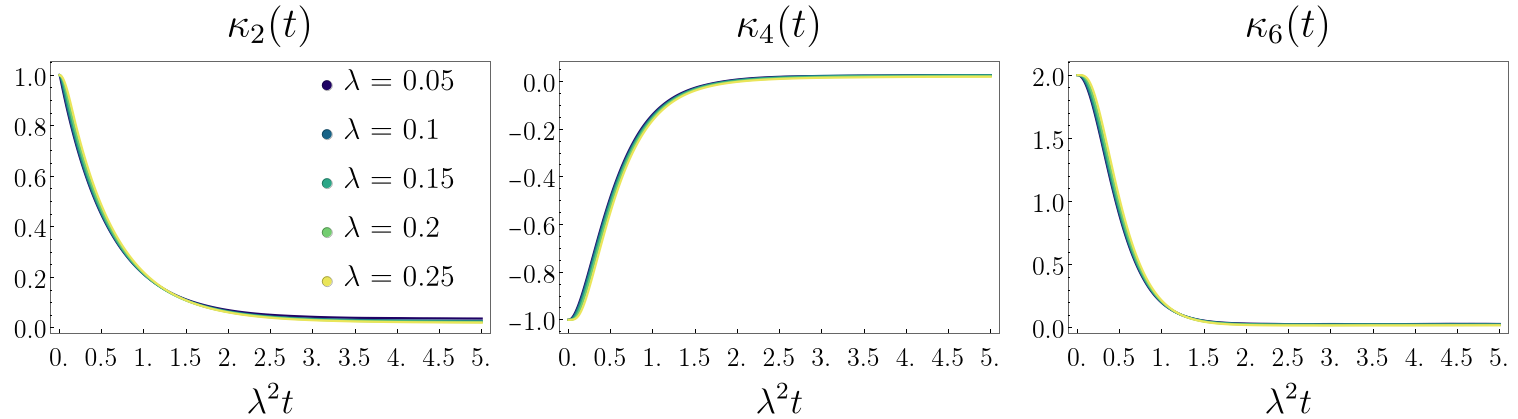}
    \caption{Thermal  free cumulants $\kappa_{n}(t)$ for $n=2,4,6$ in the random-matrix model with $L=14$ and different interaction strengths $\lambda$.}
    \label{fig:k2k4k6}
\end{figure}

In addition, we investigate specific thermal free cumulants $\kappa_{n}(t)\equiv\kappa(t,0,\ldots,t,0)$, often considered in study of the long-time freeness\cite{Vallini2024LongTime,camargo2025quantum-freeness,PhysRevB.111.014311-freeness,Fava2025Designs,PhysRevLett.129.170603-geth-Silvia}. 
The main motivation to study this specific choice for the free cumulants is that their equilibration timescales are in close relation to the ``freeness timescale'' \cite{Vallini2024LongTime,free-time}.
From Eq.\ (\ref{eq_def_cum}) we infer for $n=2,4,6$ 
    \begin{align}
        \kappa_2(t)&=\langle{A}(t){A}\rangle,\\
    \kappa_4(t)&=\langle{A}(t){A}\ {A}(t){A}\rangle-2\kappa_2(t)^2,\\
    \kappa_6(t)&=\langle A(t)A\ A(t)A\ A(t)A\rangle-6\kappa_{2}(t)\kappa_{4}(t)-4\kappa_{3}(t)^{2}-5\kappa_{2}(t)^{3}.
\end{align}
In Fig.\ \ref{fig:k2k4k6} we depict 
the second cumulant $\kappa_2(t)$, which coincide with the autocorrelation function due to $\langle A \rangle = 0$,
as well as the higher cumulants $\kappa_4(t)$ and $\kappa_6(t)$ in the random-matrix model with size $L=14$. 
An approximate data collapse of all thermal free cumulants $\kappa_{n}(t)$ for $n = 2, 4, 6$, with respect to the rescaled time $\lambda^{2} t$, can be observed, most clearly after an initial transient time. This behavior indicates that $T_{\text{eq}}^{(n)} \propto \lambda^{-2}$, in agreement with Eq.~\eqref{eq-ET}, $T^{(n)}_{\text{eq}} \sim 2\pi / \Delta E^{(n)}_{\text{eq}}$, together with the scaling $\Delta E_{\text{eq}}^{(n)} \sim \lambda^{2}$. 
In addition, Fig.~\ref{fig:kt-vs-L-rm} in the appendix shows $\kappa_n(t)$ for different $L$ at fixed coupling $\lambda = 0.2$. The near overlap of the curves suggests that $T^{(n)}_{\mathrm{eq}}$ is approximately independent of $L$, which is consistent with the approximate independence of $\Delta^{(n)}_{\mathrm{eq}}$ from $L$ observed in Fig.~\ref{fig:cum_rm_even}.
This constitutes the second main result of our paper.
Considering $T^{(n)}_{\text{eq}}$ of the thermal free cumulants $\kappa_n(t)$, from Fig.\ \ref{fig:k2k4k6} we infer that the equilibration timescales are comparable for different $n$, i.e.\ $T_{\text{eq}}^{(2)}\sim T_{\text{eq}}^{(4)}\sim T_{\text{eq}}^{(6)}$.
Since higher-order cumulants are relevant for the relaxation of far-from-equilibrium initial states, the similarity of $T_{\text{eq}}^{(n)}$suggests that the equilibration timescales for far-from-equilibrium initial states are similar to those for near-equilibrium ones.

\subsection{A chaotic Ising bath\label{subsec_ising}}
Further we investigate a quantum spin chain of similar structure. Its Hamiltonian is given by
\begin{align}
    \mathcal{H}&=\sigma_x^S+\lambda\sigma_z^S\otimes\mathcal{H}_{I}+\mathcal{H}_{B},\label{eq_ham_ising}\\
    \mathcal{H}_{I}&=\frac{1}{\sqrt{L-1}}\sum_{n=1}^{L-1}(-1)^n\sigma_z^n,\quad\\
    \mathcal{H}_{B}&=J\sum_l \sigma_z^l\sigma_z^{l+1}+h_x\sigma_x^l+h_2\sigma_z^2+h_5\sigma_z^5.
\end{align}
\begin{figure}[h]
    \centering
    \includegraphics[width=1.\linewidth]{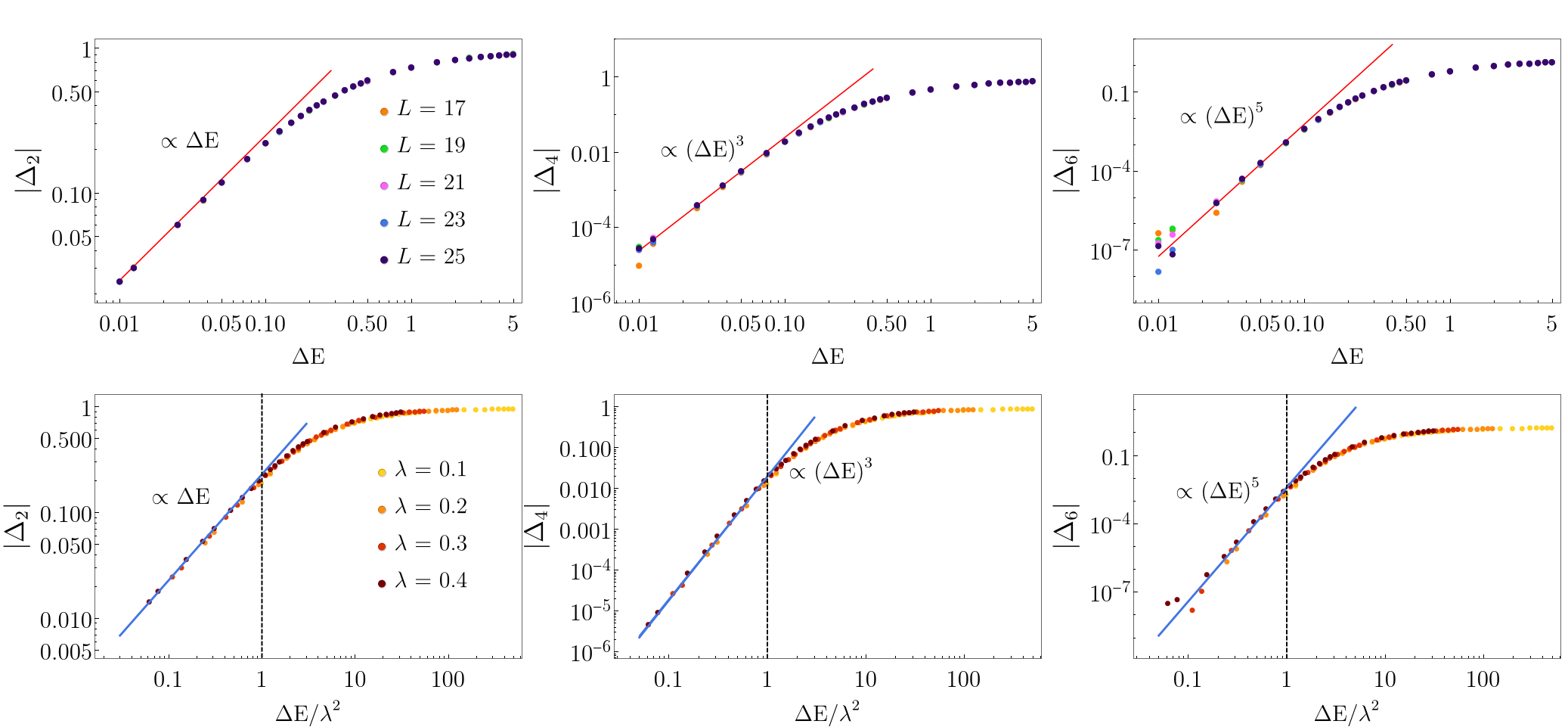}
    \caption{
    \textbf{Top:} System-size scaling of the microcanonical free cumulants $\Delta_n$ in the chaotic Ising bath model with interaction strength $\lambda=0.3$.
    \textbf{Bottom:}
    Scaling of the microcanonical free cumulants $\Delta_n$ in the chaotic Ising-bath model with system-size $L=23$ and different system-bath couplings $\lambda$. Here the dashed black line indicates $\Delta E_U$ as a guide to the eye.}
    \label{fig:cums_ising}
\end{figure}
\begin{figure}[h]
    \centering
    \includegraphics[width=1.0\linewidth]{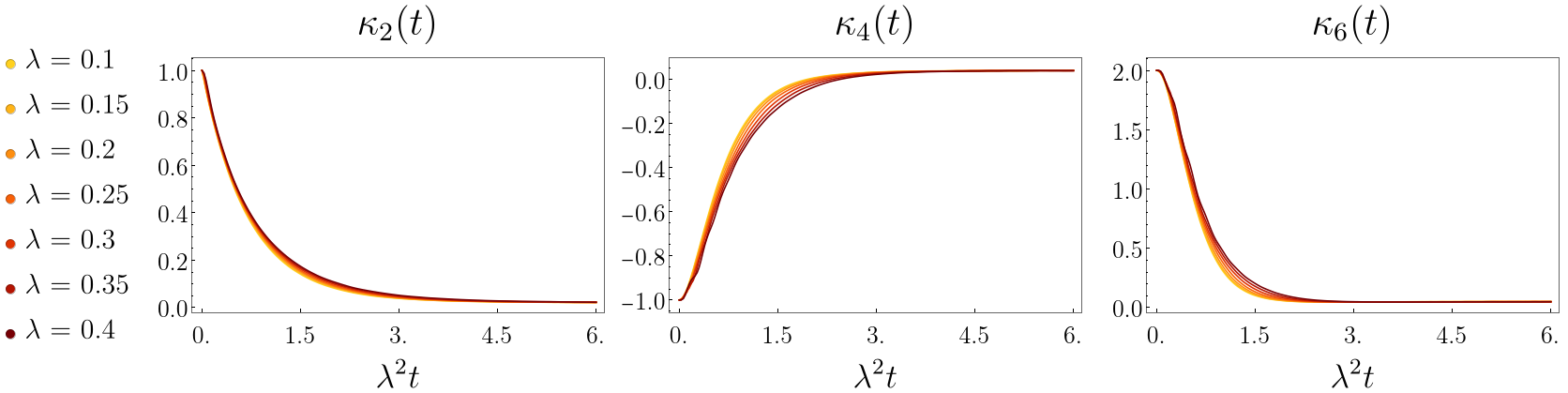}
    \caption{Thermal free cumulants $\kappa_{n}(t)$, where $n=2,4,6$, in the model with an Ising bath and total size $L=23$ and different coupling strengths $\lambda$.}
    \label{fig:ising_k2k4k6}
\end{figure}
Due to the precise form of the coupling, we restrict ourselves to odd sizes of the full system. The bath is given by a transverse Ising chain with two defects imposed at sites 2 and 5 to break symmetries and render the Hamiltonian chaotic. The parameters are chosen as $(J,h_x,h_2,h_5)=(1.0,1.0,1.11,1.61)$ and periodic boundary conditions are imposed. As with the random-matrix model above, we consider the observable ${A}=\sigma_x^S$.
Overall, similar results are observed in Figs.~\ref{fig:cums_ising}, \ref{fig:ising_k2k4k6} and \ref{fig:kt-vs-L-ising} when compared to the results for the random–matrix bath shown in Figs.~\ref{fig:cum_rm_even}, \ref{fig:k2k4k6} and \ref{fig:kt-vs-L-rm}, as detailed below.

The energy scale $\Delta E_{\text{eq}}^{(n)}$, within which Eq.~\eqref{eq_cum_powerlaw} holds, can be identified here for $n = 2, 4, 6$, and its value appears to be nearly independent of the system size (upper panel of Fig.~\ref{fig:cums_ising}).
We again observe a scaling law $\Delta_{n}(\Delta E) = f(\frac{\Delta E}{\lambda^{2}})$, which indicates the scaling $\Delta E_{\text{eq}}^{(n)}\propto \lambda^2$ (lower panel of Fig.~\ref{fig:cums_ising}).
As for the dynamics of the thermal free cumulants $\kappa_{n}(t)$ (Fig.~\ref{fig:ising_k2k4k6}), an approximate data collapse is observed as a function of $\lambda^{2}t$, although it is less pronounced than in the random–matrix bath model. This implies $T_{\text{eq}}^{(n)}\propto\lambda^{-2}$, consistent with the scaling of energy scale $\Delta E_{\text{eq}}^{(n)}\propto\lambda^{2}$.
Similar data collapse is found for different total system size $L$ (Fig.~\ref{fig:kt-vs-L-ising}), indicating the approximate independence of $T_{\text{eq}}^{(n)}$ on $L$.
Further, we again observe the equilibration times scales $T^{(n)}_{\text{eq}}$ to depend only weakly on $n$.

\section{Conclusion and Discussion\label{sec_conclusion}}

In this work, we numerically investigate the full Eigenstate Thermalization Hypothesis framework in open quantum systems. Specifically, we consider a setup 
where a central system coupled to a quantum chaotic bath. 
For observables related to the central system Hamiltonian, we find a bath-size–independent,  universal scaling of the microcanonical free cumulants with respect to the interaction strength. 
Furthermore, we established a connection between this scaling behavior and the thermalization timescales thermal free cumulants of the corresponding observables, highlighting its dynamical significance.

Since the introduction of the full ETH, most studies have focused on closed systems, while its implications for open systems remain largely unexplored. Our work represents a first step in this direction and is primarily based on numerical investigations.
The present results differ significantly from those obtained previously for closed systems.
For instance, regarding the energy scale $\Delta E^{(n)}_{\text{eq}}$, Ref.~\cite{wang-unitary-symmetry} provides numerical evidence supporting its existence. However, its value depends strongly on the system size $L$, and, in most cases, also on $n$ \cite{compare-previous}.
In contrast, in the open-system setup considered here, at least for the observable studied in the main text, $\Delta E^{(n)}_{\text{eq}}$ is largely independent of these parameters. The reason for this remains unclear at present, since a general theory analyzing higher-order cumulants in open-system settings is still lacking. Understanding this scaling behavior and its precise impact on out-of-equilibrium dynamics is an important direction for future research. Another interesting direction for future work is the study of non-Markovian dynamics, for instance in the presence of strong coupling or an integrable bath.

\section*{Acknowledgements}
This work has been funded by the Deutsche Forschungsgemeinschaft (DFG), under Grant No. 531128043, as well as under Grant No. 397107022, No. 397067869, and No. 397082825 within the DFG Research Unit FOR 2692, under Grant No. 355031190.

\section*{Data availability}
The research data associated with this article is openly available \cite{fullgraf_2026_18833341}.

\bibliography{main.bib}
\newpage
\begin{appendix}
\numberwithin{equation}{section}

\section{Details on the numerics\label{app_num_details}}
In the random-matrix model (\ref{eq_ham_random}) the computation of the microcanonical and the time-dependent cumulants $\Delta_n$ and $\kappa_n(t)$ were carried out using exact diagonalization (ED). For the microcanonical free cumulants, $N=10$ realizations were considered.

For the Ising model (\ref{eq_ham_ising}) the system sizes considered are beyond the reach of ED. Therefore both $\Delta_n$ and $\kappa_n(t)$ were obtained via techniques exploiting quantum typicality. Concretely, for the time-dependent quantities we use the concept of dynamical quantum typicality and time-evolve the respective quantities via Chebyshev-integrators. For the microcanonical free cumulants we employed the scheme suggested in Ref.\ \cite{wang-unitary-symmetry} that expands the microcanonical projector (\ref{eq_def_projector}) in terms of Chebyshev polynomials and then computes the moments (\ref{eq_def_moments}) on the basis of typicality. For more details on the method we refer to \cite{wang-unitary-symmetry}. For the expansion of the projectors we used $N_{\text{trunc}}=10a \frac{2\pi}{\Delta E}$, with $\Delta E$ the width of the energy window and $a=(E_{\text{max}}-E_{\text{min}})/2$ and $E_{\text{min}},E_{\text{max}}$ the edges of the spectrum. For every system size $L$ study, we consider $N_{\text{typ}}\ge2^{26-L}$ different realizations.
\section{Another observable in the random-matrix model\label{app_other_observable}}
\begin{figure}[h]
    \centering
\includegraphics[width=0.5\linewidth]{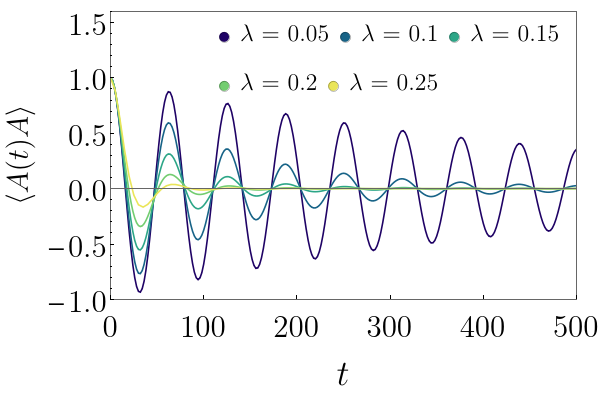}
    \caption{Autocorrelation function of the observable $A=\sigma_S^z$ in the random-matrix model (\ref{eq_ham_random}) with system size $L=14$ and different interaction strengths $\lambda$.}
    \label{fig:aat_RM_SZ}
\end{figure}

\begin{figure}[h]
    \centering
    \includegraphics[width=1\linewidth]{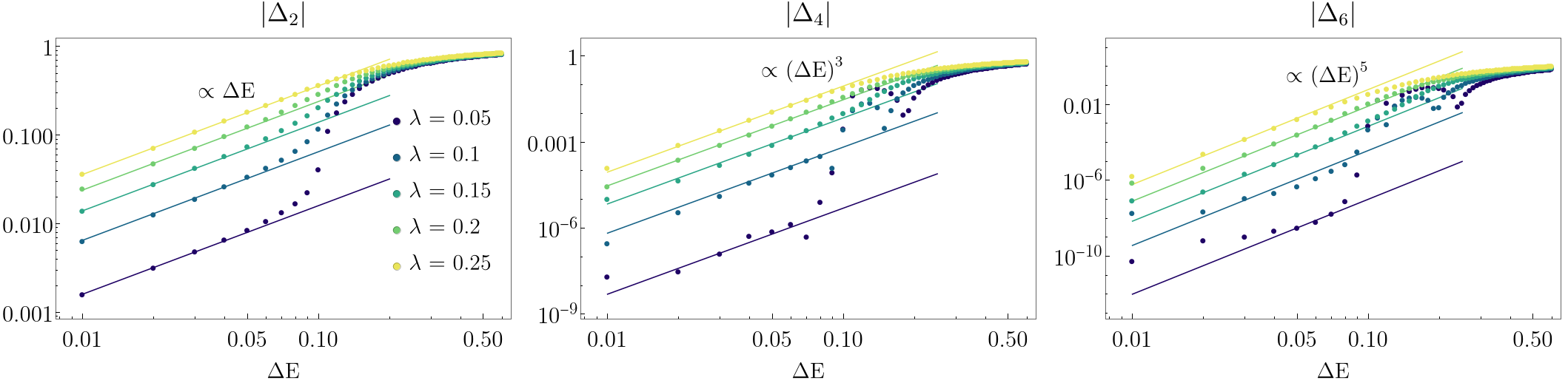}
    \caption{The first three even microcanonical free cumulants $\Delta_n$ for $A=\sigma_S^z$ in the random-matrix model. The solid lines serve as a guide to the eye for the scaling with respect to the width of the microcanonical energy window $\Delta E$, as described in Eq.\ (\ref{eq_cum_powerlaw}).}
    \label{fig:cum_RM_SZ}
\end{figure}

Here we illustrate the investigation of another observable in the random-matrix model (\ref{eq_ham_random}), given by $A=\sigma_S^z$. In Fig.\ \ref{fig:aat_RM_SZ} we depict the autocorrelation function $\langle A(t)A\rangle$, i.e.\ the second cumulant $\kappa_2(t)$ in this model. As becomes apparent in Fig.\ \ref{fig:aat_RM_SZ} $\kappa_2(t)$ does not show a coupling strength-dependent scaling with respect to time as for different $\lambda$ the overall behavior of the functions changes. The picture is similar for the microcanonical free cumulants $\Delta_n$, see Fig.\ \ref{fig:cum_RM_SZ} for the first three even cumulants. While for every interaction strength $\lambda$ the microcanonical free cumulants individually obey a power law as Eq.\ (\ref{eq_cum_powerlaw}), they do not allow for a rescaling like in the case for $A=\sigma_S^x$ expanded on in the main text.
\newpage
\section{Finite-size scalings of the free cumulants}
\begin{figure}[h]
    \centering
    \includegraphics[width=0.95\linewidth]{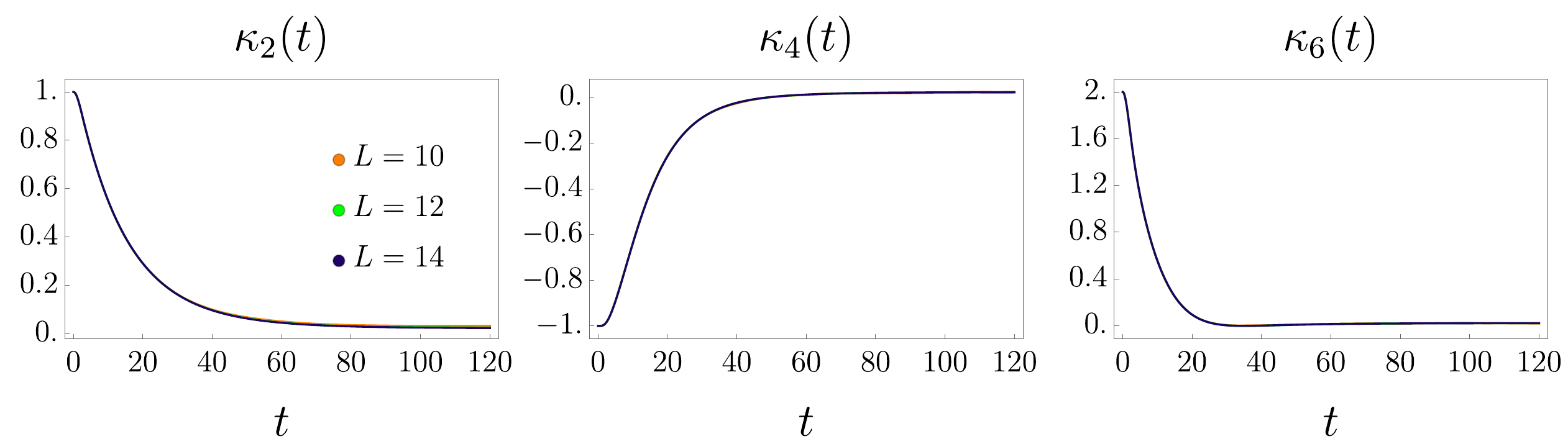}
    \caption{Thermal free cumulants $\kappa_{n}(t)$ for $n=2,4,6$ in the random-matrix model with fixed interaction strength $\lambda = 0.2$ and different $L = 10,12,14$.}
    \label{fig:kt-vs-L-rm}
\end{figure}
\begin{figure}[h]
    \centering
    \includegraphics[width=0.95\linewidth]{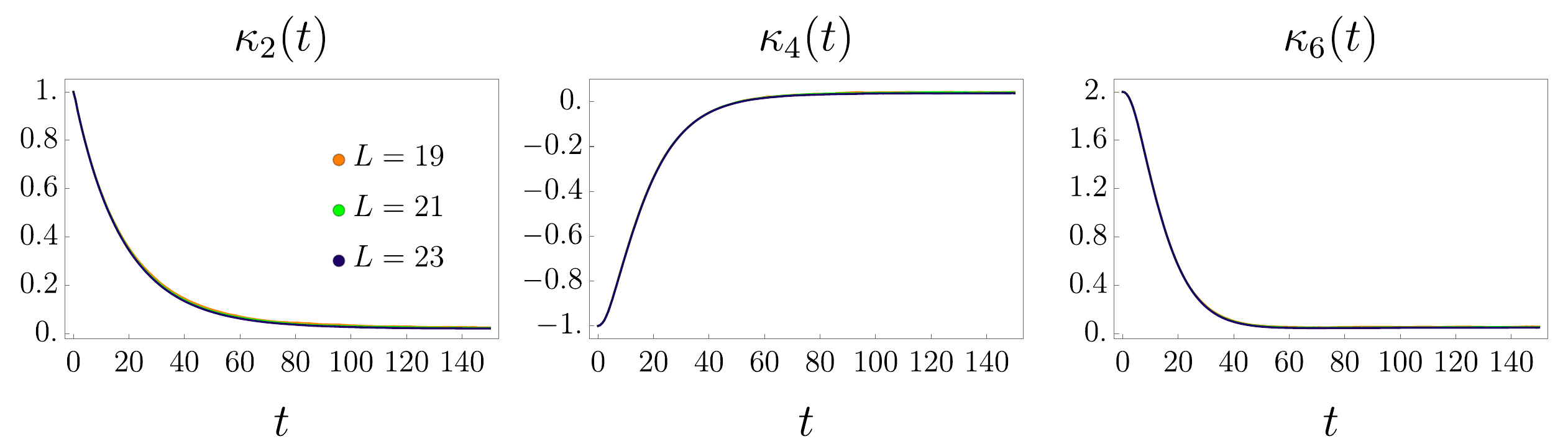}
    \caption{Thermal free cumulants $\kappa_{n}(t)$, where $n=2,4,6$, in the model with an Ising bath with fixed interaction strength $\lambda = 0.2$ and different total system sizes $L = 19,21,23$.}
    \label{fig:kt-vs-L-ising}
\end{figure}

\end{appendix}

\end{document}